\documentclass[3p,times,procedia,sort&compress]{elsarticle}

\usepackage{nupha_ecrc}

\usepackage[utf8]{inputenc}
\usepackage{amsmath}
\usepackage{wrapfig}
\usepackage{xcolor}

\graphicspath{{fig/}}

\volume{00}
\firstpage{1}
\journalname{Nuclear Physics A}
\runauth{J. Scott Moreland et al.}
\jid{nupha}
\jnltitlelogo{Nuclear Physics A}

\newcommand{\trento}{T\raisebox{-0.3ex}{R}ENTo}
\newcommand{\sqrts}{\sqrt{s_\mathrm{NN}}}
\newcommand{\T}{\tilde{T}}

\begin{document}

\begin{frontmatter}

\dochead{XXVIIth International Conference on Ultrarelativistic Nucleus-Nucleus Collisions\\ (Quark Matter 2018)}
% set empty when submitting to NPA
%\dochead{}

\title{Estimating nucleon substructure properties in a\\unified model of p-Pb and Pb-Pb collisions}

\author{J. Scott Moreland}
\ead{jsm55@phy.duke.edu}
\author{Jonah E. Bernhard}
\author{Steffen A. Bass}

\address{Department of Physics, Duke University, Durham, NC 27708-0305}

\begin{abstract}
  We apply a well tested hybrid transport model, which couples viscous hydrodynamics to a hadronic afterburner, to describe bulk observables in proton-lead and lead-lead collisions at \mbox{$\sqrts=5.02$~TeV}.
  The quark-gluon plasma (QGP) initial conditions are modeled using the parametric \trento\ model with additional nucleon substructure parameters to vary the number and size of hot spots inside each nucleon, followed by a pre-equilibrium free streaming stage to match the full energy-momentum tensor of the initial state onto viscous hydrodynamics.
  Initial condition and QGP medium parameters, such as the temperature dependence of the QGP shear and bulk viscosities, are then calibrated using Bayesian parameter estimation to describe charged particle yields, mean $p_T$ and anisotropic flow harmonics of both collision systems in a single self-consistent framework.
  We find that the hybrid model provides a compelling, simultaneous description of both collision systems using appropriately chosen model parameters, and present new posterior estimates for the size and shape of the nucleon and temperature dependence of QGP shear and bulk viscosities.
\end{abstract}


\end{frontmatter}

\section{Introduction}

Recently, long-range multiparticle correlations have been observed in high-multiplicity p-Pb collisions which are strikingly similar to correlations observed in Pb-Pb collisions and commonly attributed to hydrodynamic flow.
The strongest justification for hydrodynamic behavior in Pb-Pb collisions is the global, self-consistent, and highly nontrivial agreement of hydrodynamic simulations with experiment.
It is thus natural to wonder if similarly accurate quantitative descriptions of p-Pb bulk observables can be obtained from the hydrodynamic standard model.
Perhaps the most important constraint for such a framework, is that it describe p-Pb and Pb-Pb collision data \emph{simultaneously} within the appropriate limits of applicability, using a single set of model parameters and without additional fine tuning.
In this work we investigate the existence of such a description, and calibrate a successful hybrid model---based on viscous hydrodynamics and Boltzmann transport---to simultaneously describe p-Pb and Pb-Pb bulk observables at $\sqrts=5.02$~TeV.

\section{Physics model}

We simulate the QGP initial conditions using a modified version of the \trento\ model \cite{Moreland:2014oya} which includes parametric nucleon substructure.
Consider the collision of two protons $A, B$ with three-dimensional densities $\rho_{A,B}$ defined as a superposition of Gaussian constituents
\begin{equation}
  \label{density}
  \rho_{A,B}(\textbf{x}) = \sum\limits_{i=1}^{M} \frac{1}{(2 \pi w^2)^{3/2}} \exp\left(-\frac{(\textbf{x}-\textbf{x}_i)^2}{2 w^2}\right),
\end{equation}
where $M$ is the number of constituents and $w$ their Gaussian width.
Here the proton's constituent positions $\textbf{x}_i$ are sampled from a Gaussian radial distribution of width $\sigma = \sqrt{r^2 - w^2}$, where free parameter $r$ varies the ensemble-averaged proton radius, and $w$ is the constituent width in Eqn.~\eqref{density}.

The two protons collide inelastically at impact parameter $b$ with probability
\begin{equation}
  \label{pcoll}
  P_\text{coll}(b) = 1 - \exp\left[-\sigma_\text{eff} \int dx\,dy \int dz\, \rho_A(\mathbf{x}) \int dz\, \rho_B(\mathbf{x} + \mathbf{b}) \right],
\end{equation}
where $\sigma_\text{eff}$ is a nuisance parameter tuned to fit the total proton-proton inelastic cross section at \mbox{$\sqrts=5.02$~TeV}.
Assuming the protons collide, each is assigned a \emph{fluctuated} participant thickness function
\begin{equation}
  \label{part}
  \T_{A,B}(\mathbf{x}) = \sum\limits_{i=1}^{M} \frac{\gamma_i}{2 \pi w_c^2} \exp\left(-\frac{(\textbf{x}-\textbf{x}_i)^2}{2 w_c^2}\right),
\end{equation}
which integrates out the beam-axis ($\hat{z}$) dimension of Eqn.~\eqref{density} and assigns additional random weights $\gamma_i$ to each constituent sampled from a Gamma distribution with unit mean and variance $1/k^2$.

\begin{figure}
  \includegraphics[width=.5\textwidth]{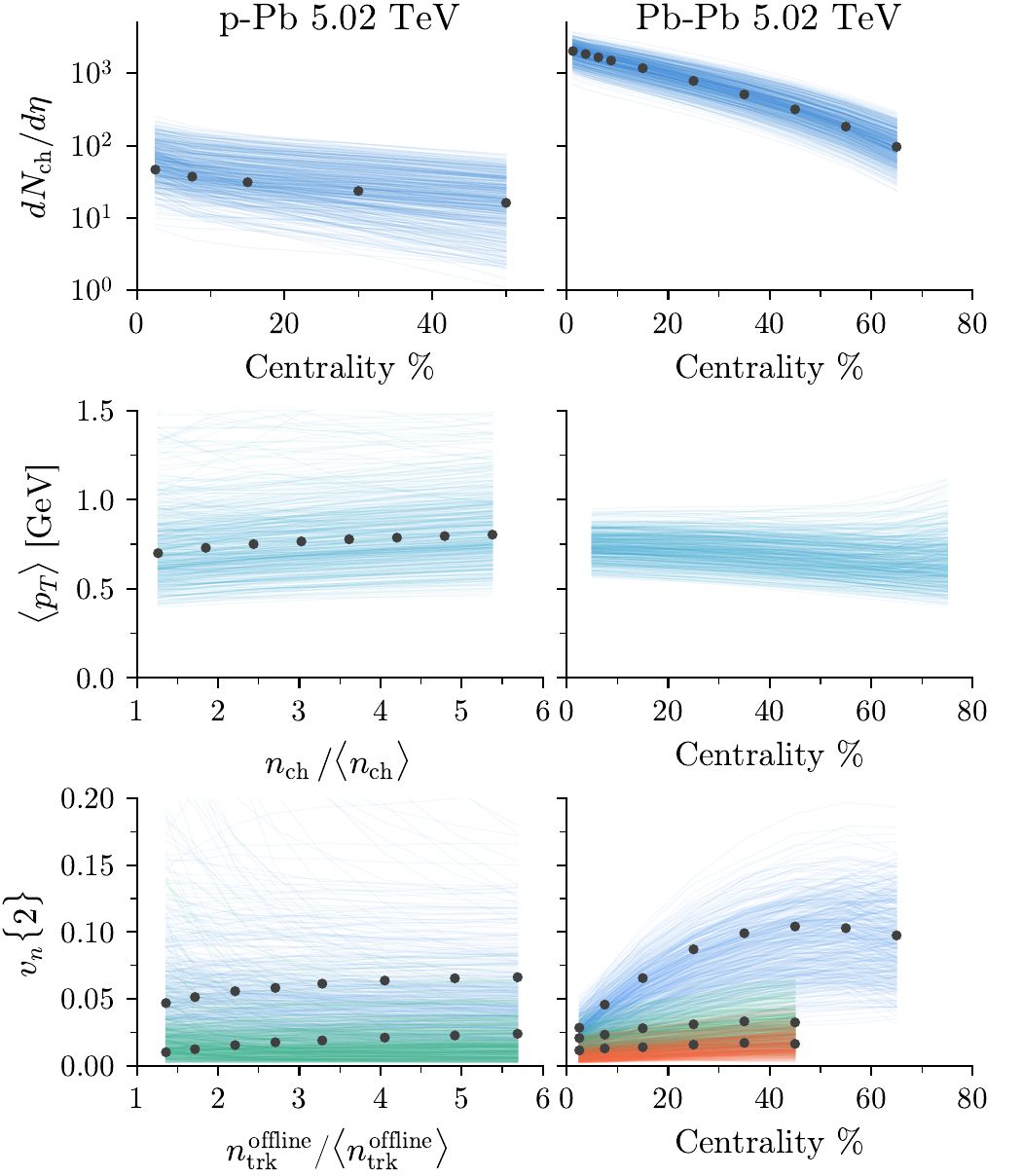}
  \includegraphics[width=.5\textwidth]{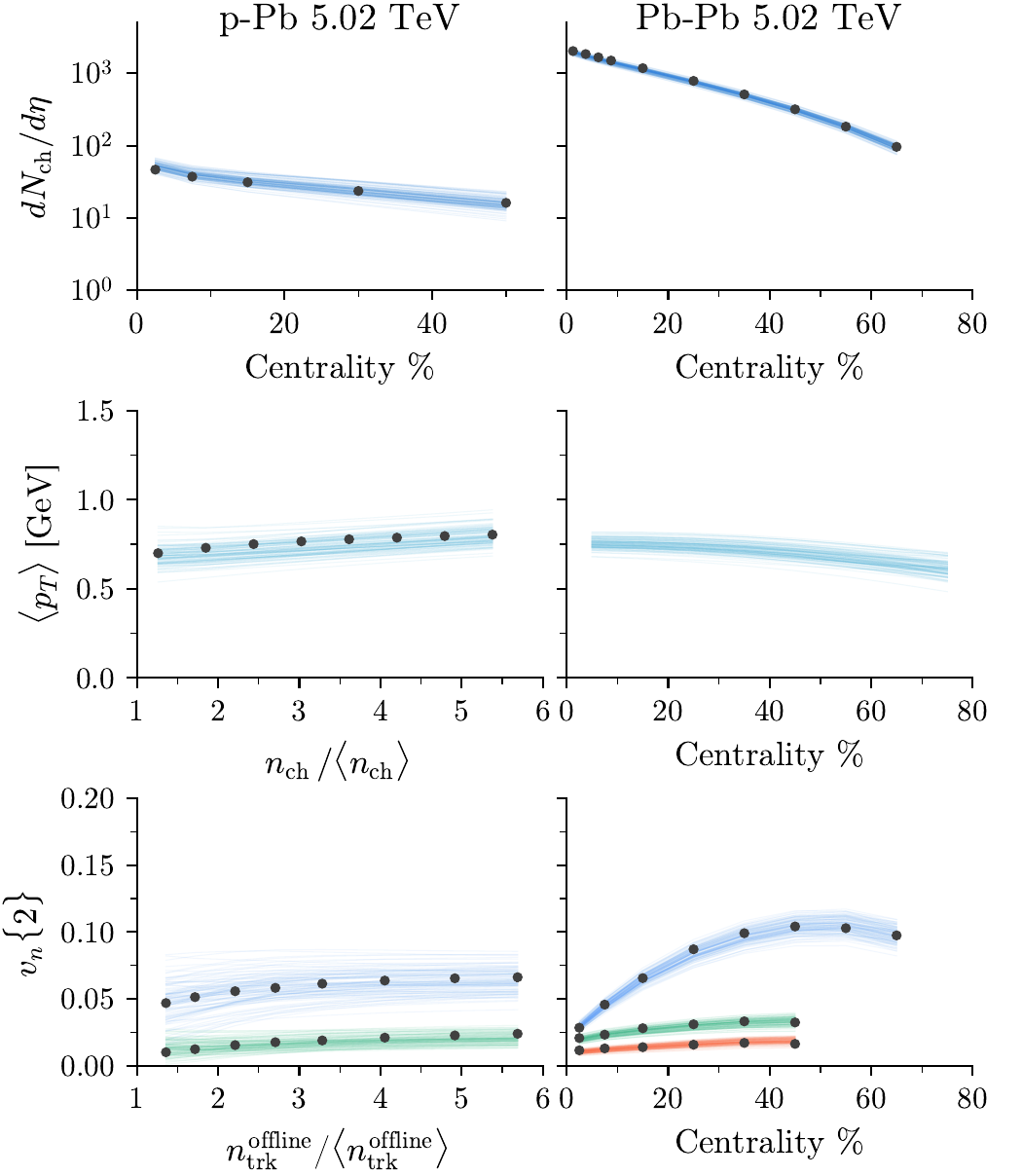}
  \caption{\label{fig:calibration} Left-side: computed observables at each design point. Right-side: posterior samples from the calibrated emulator.}
\end{figure}

The \trento\ model then deposits entropy in the proton-proton collision proportional to the generalized mean of participant matter
\begin{equation}
  \label{gmean}
  \frac{dS}{d\eta} \propto \left(\frac{\T_A^p + \T_B^p}{2} \right)^{1/p},
\end{equation}
where the dimensionless parameter ${-\infty \le p \le \infty}$ modulates the scaling behavior of initial entropy deposition.
The model generalizes readily to proton-lead and lead-lead collisions by calculating the participation probability $P_\text{coll}$ in Eqn.~\eqref{pcoll} for all pairs of colliding nucleons and summing the fluctuated participant thickness function $\T_{A,B}$ in Eqn.~\eqref{part} over all participants.

We use Eqn.~\eqref{gmean} to generate boost invariant initial conditions for p-Pb and Pb-Pb collisions at $\sqrts=5.02$~TeV, and free stream the initial profiles for proper time $\tau_{fs}$ \cite{Liu:2015nwa}.
The full energy-momentum tensor is then matched to the \texttt{VISH2+1} hydrodynamics code \cite{Song:2007ux, Shen:2014vra} which includes shear and bulk viscous corrections in the 14-moment approximation \cite{Bernhard:2016tnd}.
The hydrodynamic evolution is subsequently matched to a microscopic Boltzmann transport model along a pre-specified switching isotherm $T_\text{sw}$ treated as a free parameter.
We sample particles from the switching isotherm with shear and bulk corrections \cite{Bernhard:2018hnz}, and simulate subsequent hadronic interactions using the \texttt{UrQMD} microscopic transport model \cite{Bass:1998ca}.
Finally, we calculate hadronic observables on the particles using the same kinematic cuts and methods used by experiment.

\begin{wrapfigure}[25]{r}{.45\textwidth}
  \includegraphics[width=.45\textwidth]{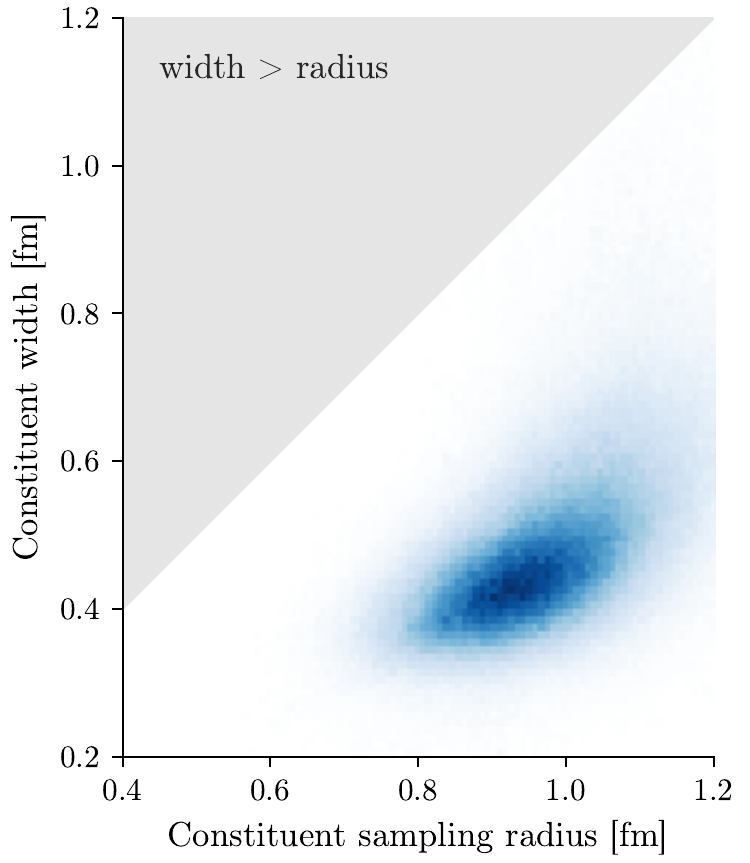}
  \caption{\label{fig:proton_shape} Joint posterior probability distribution for the constituent sampling radius $r$ and constituent width $w$. The gray shaded region $w > r$ is excluded by the chosen prior.}
\end{wrapfigure}

\section{Model calibration}

The parameter space of the present hybrid model is explored using Bayesian parameter estimation methods previously adapted and applied to heavy-ion collisions.
We use a Latin hypercube algorithm to generate a small, semi-random sample of parameter values $X$, commonly referred to as a \emph{design}, which covers a conservative initial range of values in each parameter dimension.
We then run the full hybrid model at each set of parameter values $\mathbf{x} \in X$ and calculate the midrapidity charged particle yield $dN_\text{ch}/d\eta$, mean $p_T$, and flow harmonics $v_2$, $v_3$ and $v_4$ for each collision system at $\sqrts=5.02$~TeV.
These observables are concatenated into a single observable vector $\mathbf{y}$ for each design point $\mathbf{x}$ to form the computed observables array $Y$.

The posterior probability for the true model parameters $\mathbf{x_\star}$ given the design $X$, computed observables $Y$, and experimental data $\mathbf{y}_\mathrm{exp}$ is provided by Bayes' theorem
\begin{equation}
  \label{posterior}
  P(\mathbf{x_\star}| X, Y, \mathbf{y}_\mathrm{exp}) \propto P(X, Y, \mathbf{y}_\mathrm{exp} | \mathbf{x_\star}) P(\mathbf{x_\star}).
\end{equation}
The term $P(\mathbf{x_\star})$ on the right-side is the \emph{prior} probability which encapsulates initial uncertainty on $\mathbf{x_\star}$, while $P(X, Y, \mathbf{y}_\mathrm{exp} | \mathbf{x_\star})$ is the likelihood function, defined as the probability of observing ($X$, $Y$, $\mathbf{y}_\text{exp}$) given a proposal for $\mathbf{x}_\star$:
\begin{equation}
  P(X, Y, \mathbf{y}_\mathrm{exp} | \mathbf{x_\star}) \propto \exp \left[ -\frac{1}{2}(\mathbf{y}(\mathbf{x_\star}) - \mathbf{y}_\mathrm{exp})^\top \Sigma_y^{-1} (\mathbf{y}(\mathbf{x_\star}) - \mathbf{y}_\mathrm{exp}) \right].
\end{equation}
Here $\Sigma_y$ on the right-side is the covariance matrix which combines various sources of experimental, model and emulator uncertainty.
\begin{figure}
  \centering
  \includegraphics[width=.9\textwidth]{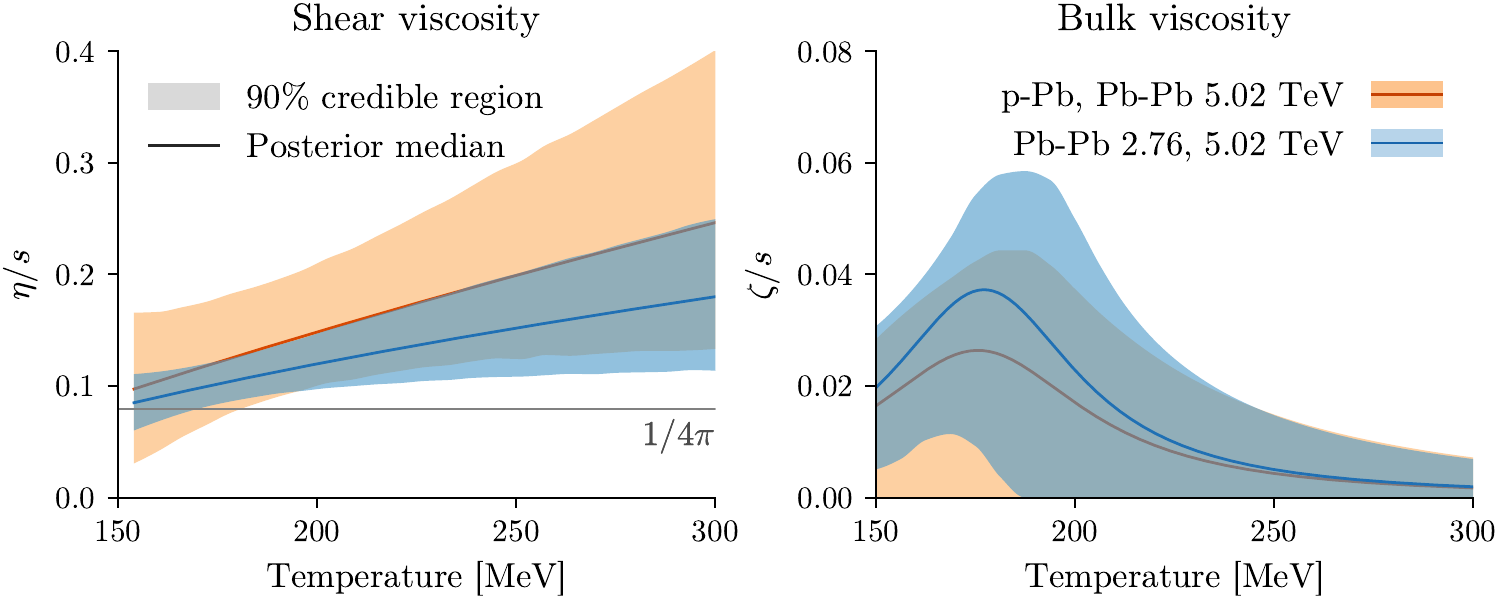}
  \caption{Posterior 90\% credible regions for the temperature dependent shear viscosity (left) and temperature dependent bulk viscosity (right) from the present combined p-Pb and Pb-Pb analysis at $\sqrts=5.02$~TeV (orange band), compared to a previous Bayesian analysis of Pb-Pb collisions at $\sqrts=2.76$ and $5.02$~TeV.}
  \label{fig:viscosity}
\end{figure}
We then perform Markov-chain Monte Carlo (MCMC) importance sampling on the posterior probability in Eqn.~\eqref{posterior}, and histogram the MCMC samples to visualize the marginalized posterior distribution of specific model parameters.

\section{Results}

Figure~\ref{fig:calibration} shows the computed model observables at each design point (left) alongside emulator predictions for $n=100$ parameter samples (right) drawn from the calibrated posterior.
We observe good agreement between the calibrated model and experiment, as illustrated by the clustering of the posterior emulator predictions around the experimental data (black points).

Figure~\ref{fig:proton_shape} shows the posterior distribution on the constituent sampling radius $r$ and constituent width $w$.
The lower-white trapezoidal region is our chosen prior and was constructed to exclude the upper-gray triangular region $w > r$ which samples constituent widths larger than our target proton width.
The posterior distribution, shown by the blue ellipsoidal region, is highly constrained and prefers a rather large sampling radius $r=0.97_{-0.16}^{+0.19}$~fm and large constituent width $w=0.47_{-0.16}^{+0.22}$~fm.
%It's important to note that our sampling radius $r$ is \emph{not} equivalent to the proton's center of mass root-mean-square radius $\sqrt{\langle r^2 \rangle}$ often quoted when discussing the proton size, although differences between the two are small when the number of constituents is large.
The marginalized posterior on the constituent number (not shown) disfavors one constituent (no substructure), but has no strong preference for a specific number of constituents $M>1$.

In Fig.~\ref{fig:viscosity} we compare the posterior median and 90\% credible regions for the temperature dependence of the QGP shear and bulk viscosities which were parametrized using the functional forms detailed in Refs~\cite{Bass:2017zyn, Bernhard:2018hnz}.
The orange line/band are obtained from the present analysis of p-Pb and Pb-Pb data at $\sqrts=5.02$~TeV, while the blue line/band are from an independent analysis of Pb-Pb data at two different beam energies $\sqrts=2.76$ and $5.02$~TeV \cite{Bernhard:2018hnz}.
We find that our estimates for the shear viscosity are fully consistent between the two studies, with the present p-Pb and Pb-Pb analysis providing significantly less constraining power than the Pb-Pb analysis at $\sqrts=2.76$ and $5.02$~TeV.
This is not surprising, given the importance of multiple beam energies to constraining temperature dependent quantities and the stronger expected effect of viscosity on larger, longer-lived collision systems.
Similarly, the 90\% credible regions on the QGP bulk viscosity are also consistent between the two studies, with both analyses favoring smaller values of $\zeta/s$ relative to $\eta/s$.
Moreover, the present study appears to provide a similar degree of constraining power on $\zeta/s$ which suggests that small-systems may play an important role in understanding the radial expansion of the produced QGP medium.

\vspace{.5em}\noindent\textbf{Acknowledgements}:
The Duke QCD group acknowledges support by grants no.\ NSF-ACI-1550225 (NSF) and DE-FG02-05ER41367 (DOE).
CPU time was provided by the DOE funded National Energy Research Scientific Computing Center (NERSC).

\bibliographystyle{elsarticle-num}
\bibliography{proceedings}

\end{document}